\documentstyle[preprint]{ptptex}
\newcommand{\ket}[1]{| {#1} \rangle}     %%
\newcommand{\kket}[1]{| {#1} \rangle\!\rangle}     %%
\title{%        %You can use \\ for explicit line-break
Note on the Orthogonal Set in\\
Six Kinds of Boson Operators
}
\subtitle{
In Relation to the su(1,1)- and its Relevant Algebras
}    %use this when you want a subtitle

\author{%       %Use \sc for the family name
Atsushi {\sc Kuriyama},$^{1}$ 
Constan\c{c}a {\sc Provid\^encia},$^{2}$ \\
Jo\~ao da {\sc Provid\^encia},$^{2}$ Yasuhiko {\sc Tsue}$^{3}$ 
and Masatoshi {\sc Yamamura}$^{1}$
}

\inst{%         %Affiliation, neglected when [addenda] or [errata]
$^1$Faculty of Engineering, Kansai University, Suita 564-8680, Japan\\
$^{2}$Departamento de Fisica, Universidade de Coimbra, P-3000 Coimbra, 
Portugal\\
$^{3}$Physics Division, Faculty of Science, Kochi University, Kochi 780-8520, 
Japan
}

\recdate{%      %Editorial Office will fill in this.
}

\abst{%       %this abstract is neglected when [addenda] or [errata]
Following the basic idea proposed by the present authors in recent paper, 
a possible form of the orthogonal set for many-body system consisting of 
six kinds of boson operators is developed. In contrast to the recent paper, 
in which the $su(2)$-algebra plays a central role, in this paper, 
the $su(1,1)$-algebra is in a central position. The orthogonal set 
obtained in this paper is expressed in terms of monomial with respect to 
the state generating operators. 
}

\begin{document}

\maketitle

%\section{Introduction}

In recent paper referred to as (I),\cite{1} we presented a possible 
form of the orthogonal set for description of many-body systems in six kinds 
of bosons. The reason why such orthogonal set is necessary was already 
mentioned in (I) and, then, in this paper, we do not repeat the reason. 
Its basic idea comes from the paper by the present authors,\cite{2} 
in which the Schwinger boson representation of an extended 
$(M+1)(N+1)$-dimensional algebra containing the $su(M+1)$- and the 
$su(N,1)$-algebras was formulated. In (I), the case $(M=1, N=2)$ was 
treated and the $su(2)$-algebra in six kinds of bosons is the object 
of the investigation. In this note, we treat the case $(M=2, N=1)$, in 
which the $su(1,1)$-algebra in six kinds of bosons can be formulated. 
Combining with the deformed boson scheme proposed by the present 
authors,\cite{3} both cases may be helpful for analyzing more realistic 
systems than those treated in Ref. \citen{3}.

First, let us recapitulate the $su(1,1)$-algebra in six kinds of bosons 
in a form slightly different from that developed by the present 
authors in Ref. \citen{2} including the notations. In the same notations as 
those in (I), the six kinds of bosons are denoted as 
$({\hat a}_i , {\hat a}_i^*)$ and $({\hat b}_i , {\hat b}_i^*)$ ($i=1,2,3$). 
In terms of these bosons, the following operators are defined : 
\begin{equation}\label{1}
{\hat T}_{\pm,0}=\sum_{i=1}^3 {\hat T}_{\pm,0}(i) \ , 
\end{equation}
\vspace{-0.2cm}
\begin{subequations}\label{2}
\begin{equation}\label{2a}
{\hat T}_+(i)={\hat b}_i^*{\hat a}_i^* \ , \qquad
{\hat T}_-(i)={\hat a}_i{\hat b}_i \ , \qquad
{\hat T}_0(i)=(1/2)\cdot ({\hat b}_i^*{\hat b}_i+{\hat a}_i^*{\hat a}_i
+1) \ .
\end{equation}
The set $({\hat T}_{\pm,0})$ obeys the $su(1,1)$-algebra. 
Associating with the operators ${\hat T}_{\pm,0}$, we introduce the 
operators defined by 
\begin{equation}\label{2b}
{\hat T}(i)=(1/2)\cdot({\hat b}_i^*{\hat b}_i-{\hat a}_i^*{\hat a}_i+1) \ . 
\qquad (i=1,2,3)
\end{equation}
\end{subequations}
Three sets $({\hat T}_{\pm,0}(i) \ ; \ i=1,2,3)$ form the independent 
$su(1,1)$-algebras in the Schwinger boson representation\cite{4} and 
${\hat T}(i)$ denotes the magnitude of the $i$-th $su(1,1)$-spin. 
The set $({\hat T}_{\pm,0})$ in this paper plays the same role as that of 
the set $({\hat S}_{\pm,0})$, which obeys the $su(2)$-algebra, in (I). 
It should be noted that ${\hat T}(i)$ is not positive-definite.

The set $({\hat T}_{\pm,0})$ forms the addition of these three 
$su(1,1)$-spins. Main aim of this paper is to give the eigenstate of 
$({\hat {\mib T}}^2 , {\hat T}_0)$ without the limits of the conventional 
manner, i.e., successive addition of the $su(1,1)$-spins. 
Here, ${\hat {\mib T}}^2$ denotes the Casimir operator of the 
$su(1,1)$-algebra : 
\begin{equation}\label{3}
{\hat {\mib T}}^2={\hat T}_0^2-(1/2)\cdot({\hat T}_+{\hat T}_- 
+{\hat T}_-{\hat T}_+)
={\hat T}_0({\hat T}_0-1)-{\hat T}_+{\hat T}_- \ .
\end{equation}
The eigenstate of $({\hat {\mib T}}^2, {\hat T}_0)$ with the eigenvalues 
$(t(t-1) , t_0)$ satisfies the relation 
\begin{eqnarray}\label{4}
& &{\hat {\mib T}}^2\ket{(\gamma); t,t_0}=t(t-1)\ket{(\gamma); t,t_0} \ , 
\nonumber\\
& &{\hat T}_0\ket{(\gamma); t,t_0}=t_0\ket{(\gamma); t,t_0} \ . 
\end{eqnarray}
Here, $(\gamma)$ denotes a set of four quantum numbers. Solutions 
of the eigenvalue equation (\ref{4}) are classified into two groups : 
\begin{subequations}\label{5}
\begin{eqnarray}
& &t=+3/2 , \ +2,\ 5/2, \cdots, \qquad t_0=t,\ t+1,\ t+2, \cdots \ , 
\label{5a}\\
& &t=-1/2 , \ -1,\ -3/2, \cdots, \qquad t_0=1-t,\ 2-t,\ 3-t, \cdots \ . 
\label{5b}
\end{eqnarray}
\end{subequations}
In this paper, we treat the case (\ref{5a}). In this case, 
$\ket{(\gamma); t,t_0}$ can be expressed in the form 
\begin{equation}\label{6}
\ket{(\gamma); t,t_0}=({\hat T}_+)^{t_0-t}\ket{(\gamma); t} \ .
\end{equation}
Here, we call the state $\ket{(\gamma); t}$ the intrinsic state and it should 
satisfy the condition 
\begin{equation}\label{7}
{\hat T}_-\ket{(\gamma); t}=0 \ , \qquad 
{\hat T}_0\ket{(\gamma); t}=t\ket{(\gamma); t} \ .  
\end{equation} 
In this paper, we omit any numerical factor appearing in any state 
such as normalization constant.

In associating with the operators ${\hat T}_{\pm,0}$, we introduce 
two $su(3)$-algebras in the following form : 
\begin{subequations}\label{8}
\begin{eqnarray}
& &{\hat S}_+(2)={\hat a}_2^*{\hat b}_3-{\hat a}_3^*{\hat b}_2 \ , \qquad
{\hat S}_+(1)={\hat a}_1^*{\hat b}_3-{\hat a}_3^*{\hat b}_1 \ , \nonumber\\
& &{\hat S}_-(2)={\hat b}_3^*{\hat a}_2-{\hat b}_2^*{\hat a}_3 \ , \qquad
{\hat S}_-(1)={\hat b}_3^*{\hat a}_1-{\hat b}_1^*{\hat a}_3 \ , 
\label{8a}\\
& &{\hat S}_2^1={\hat a}_1^*{\hat a}_2-{\hat b}_2^*{\hat b}_1 \ , \qquad
{\hat S}_1^2={\hat a}_2^*{\hat a}_1-{\hat b}_1^*{\hat b}_2 \ , 
\label{8b}\\
& &{\hat S}_2^2=(1/2)\cdot({\hat a}_2^*{\hat a}_2-{\hat b}_2^*{\hat b}_2)
+(1/2)\cdot({\hat a}_3^*{\hat a}_3-{\hat b}_3^*{\hat b}_3) \ , \nonumber\\
& &{\hat S}_1^1=(1/2)\cdot({\hat a}_1^*{\hat a}_1-{\hat b}_1^*{\hat b}_1)
+(1/2)\cdot({\hat a}_3^*{\hat a}_3-{\hat b}_3^*{\hat b}_3) \ , 
\label{8c}
\end{eqnarray}
\end{subequations}
\vspace{-0.5cm}
\begin{subequations}\label{9}
\begin{eqnarray}
& &{\hat R}_+(2)={\hat a}_3^*{\hat a}_2-{\hat b}_2^*{\hat b}_3 \ , \qquad
{\hat R}_+(1)={\hat b}_1^*{\hat b}_3-{\hat a}_3^*{\hat a}_1 \ , \nonumber\\
& &{\hat R}_-(2)={\hat a}_2^*{\hat a}_3-{\hat b}_3^*{\hat b}_2 \ , \qquad
{\hat R}_-(1)={\hat b}_3^*{\hat b}_1-{\hat a}_1^*{\hat a}_3 \ , 
\label{9a}\\
& &{\hat R}_2^1={\hat a}_1^*{\hat a}_2-{\hat b}_2^*{\hat b}_1 \ , \qquad
{\hat R}_1^2={\hat a}_2^*{\hat a}_1-{\hat b}_1^*{\hat b}_2 \ , 
\label{9b}\\
& &{\hat R}_2^2=(1/2)\cdot({\hat a}_2^*{\hat a}_2-{\hat b}_2^*{\hat b}_2)
-(1/2)\cdot({\hat a}_3^*{\hat a}_3-{\hat b}_3^*{\hat b}_3) \ , \nonumber\\
& &{\hat R}_1^1=(1/2)\cdot({\hat a}_1^*{\hat a}_1-{\hat b}_1^*{\hat b}_1)
-(1/2)\cdot({\hat a}_3^*{\hat a}_3-{\hat b}_3^*{\hat b}_3) \ . 
\label{9c}
\end{eqnarray}
\end{subequations}
The sets (\ref{8}) and (\ref{9}) obey the $su(3)$-algebras, respectively, 
but, they are not mutually commutable. An important relation is shown 
as follows : 
\begin{equation}\label{10}
[\ \hbox{\rm any\ of\ $({\hat T}_{\pm,0})$\ , \ 
any\ of\ the\ $su(3)$-generators\ (\ref{8})\ and\ (\ref{9})}\ ]=0 \ . 
\end{equation}

With the use of the generators (\ref{8}) and (\ref{9}), we define 
the following two sets of the operators : 
\begin{eqnarray}
{\hat M}_+&=&
+[{\hat S}_+(1) , {\hat S}_-(2) ]
=+[{\hat R}_+(2) , {\hat R}_-(1) ]
={\hat S}_2^1={\hat R}_2^1={\hat a}_1^*{\hat a}_2-{\hat b}_2^*{\hat b}_1 \ , 
\nonumber\\
{\hat M}_-&=&
-[{\hat S}_-(1) , {\hat S}_+(2) ]
=-[{\hat R}_-(2) , {\hat R}_+(1) ]
={\hat S}_1^2={\hat R}_1^2={\hat a}_2^*{\hat a}_1-{\hat b}_1^*{\hat b}_2 \ , 
\nonumber\\
{\hat M}_0&=&
{\hat S}_1^1-{\hat S}_2^2={\hat R}_1^1-{\hat R}_2^2 \nonumber\\
&=&(1/2)\cdot({\hat a}_1^*{\hat a}_1-{\hat b}_1^*{\hat b}_1)
-(1/2)\cdot({\hat a}_2^*{\hat a}_2-{\hat b}_2^*{\hat b}_2) \ , 
\label{11}\\
{\hat L}_+&=&
+[{\hat S}_+(2) , {\hat R}_-(1) ]
={\hat a}_2^*{\hat b}_1-{\hat a}_1^*{\hat b}_2 \ , 
\nonumber\\
{\hat L}_-&=&
-[{\hat S}_-(2) , {\hat R}_+(1) ]
={\hat b}_1^*{\hat a}_2-{\hat b}_2^*{\hat a}_1 \ , 
\nonumber\\
{\hat L}_0&=&
(1/2)\cdot({\hat S}_1^1+{\hat S}_2^2)+(1/2)\cdot(
{\hat R}_1^1+{\hat R}_2^2) \nonumber\\
&=&(1/2)\cdot({\hat a}_1^*{\hat a}_1-{\hat b}_1^*{\hat b}_1)
+(1/2)\cdot({\hat a}_2^*{\hat a}_2-{\hat b}_2^*{\hat b}_2) \ . 
\label{12}
\end{eqnarray}
Two sets $({\hat M}_{\pm,0})$ and $({\hat L}_{\pm,0})$ obey the 
$su(2)$-algebras independently : 
\begin{subequations}\label{13}
\begin{equation}\label{13a}
[\ \hbox{\rm any\ of\ $({\hat M}_{\pm,0})$\ , 
\ any\ of\ $({\hat L}_{\pm,0})$\ }]=0 \ . 
\end{equation}
Further, from the relation (\ref{10}) and the definitions (\ref{11}) and 
(\ref{12}), we have 
\begin{equation}\label{13b}
[\ \hbox{\rm any\ of\ $({\hat M}_{\pm,0})$\ and\ $({\hat L}_{\pm,0})$\ , 
\ any\ of\ $({\hat T}_{\pm,0})$\ }]=0 \ . 
\end{equation}
\end{subequations}
The Casimir operators ${\hat {\mib M}}^2$ and ${\hat {\mib L}}^2$ are 
given by 
\begin{eqnarray}
& &{\hat {\mib M}}^2={\hat M}_0^2+(1/2)\cdot
({\hat M}_+{\hat M}_-+{\hat M}_-{\hat M}_+) \ , 
\label{14}\\
& &{\hat {\mib L}}^2={\hat L}_0^2+(1/2)\cdot
({\hat L}_+{\hat L}_-+{\hat L}_-{\hat L}_+) \ . 
\label{15}
\end{eqnarray}
Definitions of $({\hat M}_{\pm,0})$ and $({\hat L}_{\pm,0})$ shown 
in the relations (\ref{11}) and (\ref{12}), respectively, give us
\begin{equation}\label{16}
{\hat {\mib L}}^2={\hat {\mib M}}^2 \ .
\end{equation}
Hereafter, we call the sets $({\hat M}_{\pm,0})$ and $({\hat L}_{\pm,0})$ 
the $M$- and $L$-spin, respectively.

Our next task is to investigate properties of the operators ${\hat S}_+(2)$, 
${\hat S}_+(1)$, ${\hat R}_+(1)$ and ${\hat R}_+(2)$ with respect to the 
$M$- and the $L$-spin. For this aim, the following commutation relations 
are helpful : 
\begin{eqnarray}
& &[ {\hat M}_+ , {\hat S}_+(2) ]={\hat S}_+(1) \ , \quad
[ {\hat M}_- , {\hat S}_+(2) ]=0 \ , \quad
[ {\hat M}_0 , {\hat S}_+(2) ]=-1/2\cdot {\hat S}_+(2) \ , \nonumber\\
& &[ {\hat M}_+ , {\hat S}_+(1) ]=0 \ , \quad
[ {\hat M}_- , {\hat S}_+(1) ]={\hat S}_+(2) \ , \quad
[ {\hat M}_0 , {\hat S}_+(1) ]=+1/2\cdot{\hat S}_+(1) \ , \nonumber\\
& &[ {\hat M}_+ , {\hat R}_+(1) ]={\hat R}_+(2) \ , \quad
[ {\hat M}_- , {\hat R}_+(1) ]=0 \ , \quad
[ {\hat M}_0 , {\hat R}_+(1) ]=-1/2\cdot {\hat R}_+(1) \ , \nonumber\\
& &[ {\hat M}_+ , {\hat R}_+(2) ]=0 \ , \quad
[ {\hat M}_- , {\hat R}_+(2) ]={\hat R}_+(1) \ , \quad
[ {\hat M}_0 , {\hat R}_+(2) ]=+1/2\cdot{\hat R}_+(2) , \quad\ 
\label{17}\\
& &[ {\hat L}_+ , {\hat R}_+(1) ]={\hat S}_+(2) \ , \quad
[ {\hat L}_- , {\hat R}_+(1) ]=0 \ , \quad
[ {\hat L}_0 , {\hat R}_+(1) ]=-1/2\cdot {\hat R}_+(1) \ , \nonumber\\
& &[ {\hat L}_+ , {\hat S}_+(2) ]=0 \ , \quad
[ {\hat L}_- , {\hat S}_+(2) ]={\hat R}_+(1) \ , \quad
[ {\hat L}_0 , {\hat S}_+(2) ]=+1/2\cdot{\hat S}_+(2) \ , \nonumber\\
& &[ {\hat L}_+ , {\hat R}_+(2) ]={\hat S}_+(1) \ , \quad
[ {\hat L}_- , {\hat R}_+(2) ]=0 \ , \quad
[ {\hat L}_0 , {\hat R}_+(2) ]=-1/2\cdot {\hat R}_+(2) \ , \nonumber\\
& &[ {\hat L}_+ , {\hat S}_+(1) ]=0 \ , \quad
[ {\hat L}_- , {\hat S}_+(1) ]={\hat R}_+(2) \ , \quad
[ {\hat L}_0 , {\hat S}_+(1) ]=+1/2\cdot{\hat S}_+(1) \ . 
\label{18}
\end{eqnarray}
The relation (\ref{17}) tells us that the sets 
$({\hat S}_+(2), {\hat S}_+(1))$ and $({\hat R}_+(1), {\hat R}_+(2))$ 
are spherical tensors with rank 1/2, respectively, for the $M$-spin. 
In the same meaning as that of the above, the relation (\ref{18}) shows 
that the sets $({\hat R}_+(1) , {\hat S}_+(2))$ and 
$({\hat R}_+(2) , {\hat S}_+(1))$ are spherical tensors with rank 1/2, 
respectively, for the $L$-spin. As is later shown in the relation (\ref{23}), 
this point is useful for constructing the intrinsic state $\ket{(\gamma) ,t}$. 
Further, we note that the operator ${\hat P}_+$ defined in the following is 
a scalar with respect to both the $M$- and the $L$-spins : 
\begin{eqnarray}\label{19}
{\hat P}_+&=&(1/2)\cdot[{\hat S}_+(2){\hat R}_+(2)
-{\hat S}_+(1){\hat R}_+(1)] \nonumber\\
&=&{\hat a}_3^*{\hat b}_3[1+(1/2)\cdot({\hat a}_1^*{\hat a}_1
+{\hat b}_1^*{\hat b}_1)+(1/2)\cdot({\hat a}_2^*{\hat a}_2
+{\hat b}_2^*{\hat b}_2)] \nonumber\\
& & -({\hat b}_1^*{\hat a}_1^*+{\hat b}_2^*{\hat a}_2^*)\cdot(1/2)\cdot
{\hat b}_3^{*2}-(1/2)\cdot{\hat a}_3^{*2}\cdot({\hat a}_1{\hat b}_1
+{\hat a}_2{\hat b}_2) \ .
\end{eqnarray}
The operator ${\hat P}_+$ commutes with ${\hat S}_+(2)$, ${\hat S}_+(1)$, 
${\hat R}_+(2)$ and ${\hat R}_+(1)$.

Finally, we contact with the Casimir operator, which we denote as 
${\hat {\mib S}}^2$, for the $su(3)$-algebra defined in the relation 
(\ref{8}). The operator ${\hat {\mib S}}^2$ is given in the following 
form : 
\begin{eqnarray}\label{20}
{\hat {\mib S}}^2&=&
(4/3)\cdot[({\hat S}_1^1)^2-{\hat S}_1^1{\hat S}_2^2+({\hat S}_2^2)^2] 
+(1/2)\cdot({\hat S}_2^1{\hat S}_1^2+{\hat S}_1^2{\hat S}_2^1) \nonumber\\
& &+(1/2)\cdot[{\hat S}_+(1){\hat S}_-(1)+{\hat S}_-(1){\hat S}_+(1)]
+(1/2)\cdot[{\hat S}_+(2){\hat S}_-(2)+{\hat S}_-(2){\hat S}_+(2)] \ . 
\nonumber\\
\end{eqnarray}
The above expression can be rewritten as 
\begin{equation}\label{21}
{\hat {\mib S}}^2=(1/3)\cdot
{\hat S}_0({\hat S}_0-3)+{\hat {\mib M}}^2+
[{\hat S}_+(1){\hat S}_-(1)+{\hat S}_+(2){\hat S}_-(2)] \ .
\end{equation}
Here, ${\hat S}_0$ is defined in the form 
\begin{eqnarray}\label{22}
{\hat S}_0&=&{\hat S}_1^1+{\hat S}_2^2 \nonumber\\
&=&(1/2)\cdot({\hat a}_1^*{\hat a}_1-{\hat b}_1^*{\hat b}_1)
+(1/2)\cdot({\hat a}_2^*{\hat a}_2-{\hat b}_2^*{\hat b}_2)
+({\hat a}_3^*{\hat a}_3-{\hat b}_3^*{\hat b}_3) \ . \quad
\end{eqnarray}
The above is our framework for the later discussion. 

From the above framework, we have four mutually commutable hermitian 
operators, ${\hat {\mib M}}^2$, ${\hat M}_0$, ${\hat {\mib S}}^2$ and 
${\hat S}_0$, which are also commuted with ${\hat T}_{\pm,0}$. 
They are expressed in terms of the $su(3)$-generators defined in the 
relation (\ref{8}). The eigenvalue equations for these operators 
determine the set of four quantum numbers $(\gamma)$ appearing 
in the intrinsic state shown in the relation (\ref{6}). 
The idea presented in (I) helps us to determine $\ket{(\gamma);t}$ 
in the following form : 
\begin{subequations}\label{23}
\begin{eqnarray}
\ket{(\gamma);t}&=&
\ket{k,l,m,m_0;t} \nonumber\\
&=&\sum_{k_0l_0}C(k,k_0;l,l_0|m,m_0) {\hat S}_+(k,k_0)
{\hat R}_+(l,l_0)\cdot({\hat b}_3^*)^{2t-3} \ket{0} \ , 
\label{23x}\\
{\hat S}_+(k,k_0)&=&
\left(\sqrt{(k+k_0)!(k-k_0)!}\right)^{-1}
({\hat S}_+(1))^{k+k_0}({\hat S}_+(2))^{k-k_0} \ , 
\label{23a}\\
{\hat R}_+(l,l_0)&=&
\left(\sqrt{(l+l_0)!(l-l_0)!}\right)^{-1}
({\hat R}_+(2))^{l+l_0}({\hat R}_+(1))^{l-l_0} \ , 
\label{23b}
\end{eqnarray}
\end{subequations}
In the same meaning as that shown in (I), ${\hat S}_+(k,k_0)$ and 
${\hat R}_+(l,l_0)$ are spherical tensors with rank $k$ and $l$, respectively. 
Of course, $k_0$ and $l_0$ denote the $k_0$-th and the $l_0$-th components, 
respectively. The quantities $C(k,k_0;l,l_0|m,m_0)$ is the 
Clebsch-Gordan(CG)-coefficient. The form (\ref{23}) gives us the 
eigenvalues of ${\hat {\mib M}}^2$ and ${\hat M}_0$, 
$({\mib M}^2)_{\rm ev}$ and $(M_0)_{\rm ev}$ : 
\begin{eqnarray}\label{24}
& &({\mib M}^2)_{\rm ev}= 
m(m+1) \ , \nonumber\\
& &(M_0)_{\rm ev}=m_0 \  .
\end{eqnarray}
Further, the eigenvalue of ${\hat S}_0$, $(S_0)_{\rm ev}$, 
is obtained in the form 
\begin{equation}\label{25}
(S_0)_{\rm ev}=3k-(2t-3-l) \ .
\end{equation}
For the derivation of the above result, the following relation is useful :
\begin{eqnarray}\label{26}
& &[{\hat S}_0 , {\hat S}_+(1) ] = 3/2\cdot {\hat S}_+(1) \ , \qquad
[{\hat S}_0 , {\hat S}_+(2) ] = 3/2\cdot {\hat S}_+(2) \ , \nonumber\\
& &[{\hat S}_0 , {\hat R}_+(1) ] = 1/2\cdot {\hat R}_+(1) \ , \qquad 
[{\hat S}_0 , {\hat R}_+(2) ] = 1/2\cdot {\hat R}_+(2) \ . 
\end{eqnarray}
The eigenvalue of ${\hat {\mib S}}^2$, $({\mib S}^2)_{\rm ev}$, is given as 
\begin{equation}\label{27}
({\mib S}^2)_{\rm ev}=(1/3)\cdot(2t-3-l)(2t-l)+l(l+1) \ .
\end{equation}
The above result is obtained through the relations 
\begin{eqnarray}
& &{\hat {\mib S}}^2\ket{(\gamma);t}
=\sum_{k_0l_0}C(k,k_0;l,l_0|m,m_0){\hat S}_+(k,k_0)\cdot
{\hat {\mib S}}^2\ket{l,l_0;t} \ , 
\label{28}\\
& &\ket{l,l_0;t}={\hat R}_+(l,l_0)({\hat b}_3^*)^{2t-3}\ket{0} \ , 
\label{29}\\
& &{\hat S}_-(1)\ket{l,l_0;t}={\hat S}_-(2)\ket{l,l_0;t}=0 \ , 
\label{30}\\
& &{\hat S}_0\ket{l,l_0;t}=-(2t-3-l)\ket{l,l_0;t} \ , 
\label{31}\\
& &{\hat {\mib M}}^2\ket{l,l_0;t}=l(l+1)\ket{l,l_0;t} \ . 
\label{32}
\end{eqnarray}
As is shown in the relation (\ref{16}), ${\hat {\mib L}}^2={\hat {\mib M}}^2$ 
and, then, the state (\ref{23}) is an eigenstate of ${\hat {\mib L}}^2$ 
with the eigenvalue $m(m+1)$. Further, the state (\ref{23}) is an eigenstate 
of ${\hat L}_0$ with the eigenvalue $(L_0)_{\rm ev}$ : 
\begin{equation}\label{33}
(L_0)_{\rm ev}=k-l \ .
\end{equation}

The state (\ref{23x}) 
is of the polynomial form and, therefore, from the reason 
mentioned in (I), it may be interesting to give the form in terms of 
the monomial. For this aim, first, we define the following state : 
\begin{equation}\label{34}
\kket{n,m;t}=({\hat P}_+)^n\ket{m,-m;t} \ .
\end{equation}
Here, ${\hat P}_+$ and $\ket{m,-m;t}$ are defined in the relations 
(\ref{19}) and (\ref{29}), respectively. The state $\kket{n,m;t}$ satisfies 
\begin{eqnarray}\label{35}
& &{\hat M}_-\kket{n,m;t}={\hat L}_-\kket{n,m;t}=0 \ , \nonumber\\
& &{\hat M}_0\kket{n,m;t}={\hat L}_0\kket{n,m;t}=-m\kket{n,m;t} \ . 
\end{eqnarray}
Next, we introduce the state 
\begin{equation}\label{36}
\kket{n,k-l,m,m_0;t}=({\hat M}_+)^{m_0+m}({\hat L}_+)^{k-l+m}
\kket{n,m;t} \ .
\end{equation}
With the help of the relation (\ref{35}), we can show that the state 
(\ref{36}) is an eigenstate of ${\hat {\mib M}}^2\ (={\hat {\mib L}}^2)$, 
${\hat M}_0$ and ${\hat L}_0$ with the eigenvalues $m(m+1)$, $m_0$ and 
$k-l$, respectively. 
These are identical with the eigenvalues shown in the relations (\ref{24}) 
and (\ref{33}), respectively. The quantum number $n$ can be determined in 
the process mentioned below. We can prove the relation 
\begin{equation}\label{37}
{\hat S}_0\kket{n,k-l,m,m_0;t}
=[(k-l)+2(m+n)-(2l-3)]\kket{n,k-l,m,m_0;t} \ .
\end{equation}
Then, equating the eigenvalue (\ref{37}) with $[3k-(2t-3-l)]$ 
shown in the relation (\ref{25}), we have 
\begin{equation}\label{38}
n=k+l-m \ .
\end{equation}
Thus, we are able to rewrite the state (\ref{23x}) in terms of the monomial 
shown in the following form : 
\begin{equation}\label{39}
\ket{k,l,m,m_0;t}
=({\hat M}_+)^{m_0+m}({\hat L}_+)^{k-l+m}({\hat P}_+)^{k+l-m}
({\hat R}_+(1))^{2m}({\hat b}_3^*)^{2t-3}\ket{0} \ .
\end{equation}
It can be proved that the state (\ref{39}) is an eigenstate of 
${\hat {\mib S}}^2$, the eigenvalue of which is given in the relation 
(\ref{27}). This state can be rewritten as 
\begin{equation}\label{40}
\ket{k,l,m,m_0;t}
=({\hat M}_+)^{m_0+m}({\hat S}_+(2))^{k-l+m}({\hat R}_+(1))^{l-k+m}
({\hat P}_+)^{k+l-m}({\hat b}_3^*)^{2t-3}\ket{0} \ .
\end{equation}
The quantities $(k-l+m)$, $(l-k+m)$ and $(k+l-m)$ in the state (\ref{40}) 
should be positive and we have 
%\begin{equation}
$$
m\ge l-k \ , \qquad m\ge k-l \ , \qquad m\le k+l \ , \nonumber
$$
%\end{equation}
that is, 
\begin{equation}\label{41}
|k-l| \le m \le k+l \ .
\end{equation}
Further, as is proved at the end of this paper, the state (\ref{39}) 
obeys the condition 
\begin{equation}\label{42}
k+l+m \le 2t-3 \ .
\end{equation}
Of course, concerning $m$, $m_0$, $t$ and $t_0$, we have the relations 
\begin{eqnarray}
& & -m \le m_0 \le +m \ , 
\label{43}\\
& &t_0 \ge t \ , \qquad t \ge 3/2 \ . 
\label{44}
\end{eqnarray}
In principle, the normalization constant can be calculated, but, practically, 
it may be not necessary to give it.

Our next problem is to investigate successive addition of three 
$su(1,1)$-spins. First, we note that there exist six mutually commutable 
hermitian operators : ${\hat {\mib T}}^2$, ${\hat T}_0$, ${\hat T}(1)$, 
${\hat T}(2)$, ${\hat T}(3)$ and ${\hat {\mib T}}_{12}^2$. 
Here, ${\hat {\mib T}}^2$, ${\hat T}_0$, ${\hat T}(1)$, ${\hat T}(2)$ 
and ${\hat T}(3)$ are given in the relations (\ref{3}), (\ref{1}) and 
(\ref{2b}), respectively. The operator ${\hat {\mib T}}_{12}^2$ denotes 
the Casimir operator of the $su(1,1)$-algebra composed of the two 
$su(1,1)$-spins specified by the indices 1 and 2 : 
\begin{eqnarray}\label{45}
{\hat {\mib T}}_{12}^2&=&
({\hat T}_0(1)+{\hat T}_0(2))^2 \nonumber\\
& &-(1/2)\!\cdot\![({\hat T}_+(1)\!+\!{\hat T}_+(2))
({\hat T}_-(1)\!+\!{\hat T}_-(2))
+({\hat T}_-(1)\!+\!{\hat T}_-(2))
({\hat T}_+(1)\!+\!{\hat T}_+(2))] \ . \nonumber\\
\end{eqnarray}
The four operators except for $({\hat {\mib T}}^2 , {\hat T}_0)$ can be 
expressed as follows :  
\begin{subequations}\label{46}
\begin{eqnarray}
& &{\hat T}(1)=-(1/2)\cdot({\hat L}_0+{\hat M}_0-1)\ , \qquad
{\hat T}(2)=-(1/2)\cdot({\hat L}_0-{\hat M}_0-1)\ , \nonumber\\
& &{\hat T}(3)=+(1/2)\cdot({\hat L}_0-{\hat S}_0+1)\ , 
\label{46a}\\
& &{\hat {\mib T}}_{12}^2={\hat {\mib M}}^2 \ (={\hat {\mib L}}^2) \ . 
\label{46b}
\end{eqnarray}
\end{subequations}
Here, for ${\hat M}_0$, ${\hat L}_0$ and ${\hat S}_0$, the relations 
(\ref{11}), (\ref{12}) and (\ref{22}) were used. The relation (\ref{46b}) 
is given by the straightforward calculation.

The state (\ref{39}) is the eigenstate of ${\hat M}_0$, ${\hat L}_0$, 
${\hat S}_0$ and ${\hat {\mib M}}^2$. Therefore, the state (\ref{39}) 
is also the eigenstate of ${\hat T}(1)$, ${\hat T}(2)$, ${\hat T}(3)$ 
and ${\hat {\mib T}}_{12}^2$, the eigenvalues of which we denote as 
$t_1$, $t_2$, $t_3$ and $t_{12}(t_{12}-1)$, respectively : 
\begin{subequations}\label{47}
\begin{eqnarray}
& &t_1=-(1/2)\cdot(k-l+m_0-1) \ , \qquad
t_2=-(1/2)\cdot(k-l-m_0-1) \ , \nonumber\\
& &t_3=t-1-(k+l) \ , 
\label{47a}\\
& &t_{12}(t_{12}-1)=m(m+1) \ . 
\label{47b}
\end{eqnarray}
\end{subequations}
The relation (\ref{47a}) is solved inversely : 
\begin{eqnarray}\label{48}
& &k=(1/2)\cdot(t-t_1-t_2-t_3) \ , \qquad
k+l=t-t_3-1 \ , \nonumber\\
\hbox{\rm i.e.,}& & \nonumber\\
& &l=(1/2)\cdot(t+t_1+t_2-t_3)-1 \ , \qquad 
k-l=1-(t_1+t_2) \ , \nonumber\\
& &m_0=t_2-t_1 \ . 
\end{eqnarray}
The relation (\ref{47b}) gives us the following relation :
\begin{subequations}\label{49}
\begin{eqnarray}
& &m=t_{12}-1  \ , 
\label{49a}\\
& &m=-t_{12} \ . 
\label{49b}
\end{eqnarray}
\end{subequations}
Substituting the relation (\ref{48}) and (\ref{49a}) into the inequality 
(\ref{41})$\sim$(\ref{43}), we have the inequality 
\begin{subequations}\label{50}
\begin{eqnarray}\label{50a}
& &t_{12} \ge |t_1-1/2|+|t_2-1/2|+1 \ , \nonumber\\
& &t \ge |t_{12}-1|+|t_3-1/2|+3/2 \ .
\end{eqnarray}
The relations (\ref{48}) and (\ref{49b}) also give us the inequality 
\begin{eqnarray}\label{50b}
& &t_{12} \le -(|t_1-1/2|+|t_2-1/2|) \ , \nonumber\\
& &t \ge |t_{12}|+|t_3-1/2|+3/2 \ .
\end{eqnarray}
\end{subequations}
The relations (\ref{50a}) and (\ref{50b}) show the coupling rule for 
two and three $su(1,1)$-spins. Since $m\ge 0$, we have 
$|t_{12}-1|=t_{12}-1 \ge 0$ and $|t_{12}|=-t_{12} \ge 0$. 
Substitution of the relations (\ref{48}) and (\ref{49}) into the state 
(\ref{40}) leads us to 
\begin{subequations}\label{51}
\begin{eqnarray}
\ket{t_1,t_2,(t_{12}),t_3;t}
&=&({\hat M}_+)^{(t_{12}-1)-(t_1-1/2)+(t_2-1/2)} \nonumber\\
& &\times ({\hat S}_+(2))^{(t_{12}-1)-(t_1-1/2)-(t_2-1/2)} 
({\hat R}_+(1))^{(t_{12}-1)+(t_1-1/2)+(t_2-1/2)} \nonumber\\
& &\times ({\hat P}_+)^{(t-3/2)-(t_{12}-1)-(t_3-1/2)}
({\hat b}_3^*)^{2t-3}\ket{0} \ . \quad (t_{12}-1 \ge 0) 
\label{51a}\\
\ket{t_1,t_2,(t_{12}),t_3;t}
&=&({\hat M}_+)^{(-t_{12})-(t_1-1/2)+(t_2-1/2)} \nonumber\\
& &\times ({\hat S}_+(2))^{(-t_{12})-(t_1-1/2)-(t_2-1/2)} 
({\hat R}_+(1))^{(-t_{12})+(t_1-1/2)+(t_2-1/2)} \nonumber\\
& &\times ({\hat P}_+)^{(t-3/2)-(-t_{12})-(t_3-1/2)}
({\hat b}_3^*)^{2t-3}\ket{0} \ . \quad (-t_{12} \ge 0) 
\label{51b}
\end{eqnarray}
\end{subequations}
Here, the state $\ket{t_1,t_2,(t_{12}),t_3;t}$ is relabeled from 
$\ket{k,l,m,m_0;t}$. The above is the formalism in terms of the addition of 
three $su(1,1)$-spins.

This paper closes with proving the inequality (\ref{42}). For this aim, 
we take up the following state which is related with the state (\ref{40}) : 
\begin{equation}\label{52}
\kket{s,r,p;T}=({\hat S}_+(2))^s({\hat R}_+(1))^r({\hat P}_+)^p
({\hat b}_3^*)^T\ket{0} \ .
\end{equation}
The state (\ref{40}) can be expressed as 
\begin{equation}\label{53}
\ket{k,l,m,m_0;t}=({\hat M}_+)^{m_0+m}\kket{k-l+m,l-k+m,k+l-m;2t-3} \ .
\end{equation}
The definition of ${\hat P}_+$ shown in the relation (\ref{19}) gives us the 
relation 
\begin{eqnarray}\label{54}
({\hat P}_+)^p({\hat b}_3^*)^T\ket{0}
&=&\sum_{q=0}^p D(p,q)(T!/(T-p-q)!) \nonumber\\
& &\qquad\qquad\times
({\hat b}_1^*{\hat a}_1^*+{\hat b}_2^*{\hat a}_2^*)^q({\hat a}_3^*)^{p-q}
({\hat b}_3^*)^{T-p-q} \ket{0} \ .
\end{eqnarray}
Here, $D(p,q)$ obeys the recursion formula 
\begin{equation}\label{55}
D(p,q)=-(1/2)\cdot D(p-1,q-1)+(q+1)D(p-1,q)
-(1/2)\cdot(q+1)(q+2)D(p-1,q+1) \ .
\end{equation}
Of course, we have 
\begin{equation}\label{56}
D(0,0)=1 \ , \qquad D(p,q)=0 \quad \hbox{\rm for}\quad
q\le -1 \ \ \hbox{\rm and}\ \ q\ge p+1 \ .
\end{equation}
Some examples are given in the following numbers : 
\begin{eqnarray}\label{57}
& &D(1,0)=1 \ , \qquad D(1,1)=-1/2 \ , \nonumber\\
& &D(2,0)=3/2 \ , \qquad D(2,1)=-3/2 \ , \qquad D(2,2)=1/4 \ , 
\nonumber\\
& &D(3,0)=3 \ , \quad D(3,1)=-9/2 \ , \quad D(3,2)=3/2 \ , \quad 
D(3,3)=-1/8 \ . 
\end{eqnarray}
Further, we rewrite the state (\ref{52}) under the relation 
\begin{equation}\label{58}
{\hat b}_1^*{\hat a}_1^*+{\hat b}_2^*{\hat a}_2^*
={\hat T}_+-{\hat b}_3^*{\hat a}_3^* \ , \qquad
[ {\hat T}_+ , {\hat b}_3^*{\hat a}_3^* ]=0 \ . 
\end{equation}
Then, the binomial theorem gives us 
\begin{equation}\label{59}
({\hat b}_1^*{\hat a}_1^*+{\hat b}_2^*{\hat a}_2^*)^q
=\sum_{v=0}^q (-)^{q-v}(q!/v!(q-v)!)\cdot ({\hat T}_+)^v
({\hat b}_3^*{\hat a}_3^*)^{q-v} \ .
\end{equation}
Substituting the relation (\ref{59}) into the form (\ref{54}), we have 
\begin{eqnarray}
& &({\hat P}_+)^p({\hat b}_3^*)^T\ket{0}
=\sum_{q=0}^p C(p,q)({\hat T}_+)^q({\hat a}_3^*)^{p-q}
({\hat b}_3^*)^{T-p-q} \ket{0} \ , 
\label{60}\\
& &C(p,q)=\sum_{v=0}^p D(p,v)(T!/(T-p-v)!)\cdot (-)^{v-q}
(q!/(v-q)!v!) \ .
\end{eqnarray}

Next, with the aid of the form (\ref{60}), we rewrite the state (\ref{52}). 
First, the following relation should be noted : 
\begin{equation}\label{62}
[{\hat S}_+(2) , {\hat T}_+ ]=[{\hat R}_+(1) , {\hat T}_+ ]=0 \ .
\end{equation}
Then, under the definitions of ${\hat S}_+(2)$ and ${\hat R}_+(1)$ shown 
in the relations (\ref{8a}) and (\ref{9a}), respectively, we have 
\begin{eqnarray}\label{63}
\kket{s,r,p;T}&=&
\sum_{q=0}^p C(p,q)({\hat T}_+)^q({\hat S}_+(2))^s({\hat R}_+(1))^r
({\hat a}_3^*)^{p-q}({\hat b}_3^*)^{T-p-q}\ket{0} \nonumber\\
&=&\sum_{q=0}^p C(p,q)({\hat T}_+)^q({\hat a}_2^*{\hat b}_3)^s
({\hat b}_1^*{\hat b}_3)^r({\hat a}_3^*)^{p-q}({\hat b}_3^*)^{T-p-q}
\ket{0} \nonumber\\
&=&\sum_{q=0}^p C(p,q) ((T-p-q)!/(T-p-q-s-r)!) \nonumber\\
& &\qquad\qquad
\times \cdot({\hat T}_+)^q({\hat a}_2^*)^s({\hat b}_1^*)^r({\hat a}_3^*)^{p-q}
({\hat b}_3^*)^{(T-p-s-r)-q}\ket{0} \ . 
\end{eqnarray}
For the expression (\ref{63}), let us impose the condition 
\begin{equation}\label{64}
T-p-s-r \le -1 \ .
\end{equation}
Under the condition (\ref{64}), the state $\kket{s,r,p;T}$ 
cannot exist independently of any other condition. 
Therefore, we can conclude that the condition supporting the 
existence of the state $\kket{s,r,p;T}$ is expressed as 
\begin{equation}\label{65}
T-p-s-r \ge 0 \ .
\end{equation}
If $s=k-l+m$, $r=l-k+m$, $p=k+l-m$ and $T=2t-3$, the condition (\ref{65}) 
is rewritten in the form 
\begin{equation}\label{66}
k+l+m \le 2t-3 \ .
\end{equation}
The condition (\ref{66}) is identical with the inequality (\ref{42}).

Thus, we can conclude that the orthogonal set in six kinds of 
bosons can be formulated in relation to the $su(1,1)$-algebra 
in six kinds of bosons.

%\section*{Acknowledgements}

\end{document}